\documentclass[aps,prl,reprint]{revtex4-1}
\usepackage{setspace}
\usepackage{amsfonts,amsmath,amssymb}
\usepackage{txfonts}
\usepackage{graphicx}
\usepackage{bm}
\usepackage{verbatim}
\usepackage{hyperref}
\usepackage{units}
\usepackage[T1]{fontenc}
\usepackage[latin9]{inputenc}
\usepackage{amsbsy}
\usepackage{esint}

\begin{document}

\title{Generation of large-scale magnetic fields by small-scale dynamo in
shear flows}

\author{J.~Squire}
\email{jsquire@princeton.edu}
%\affiliation{Department of Astrophysical Sciences and Princeton Plasma Physics Laboratory, Princeton University,Princeton, NJ 08543}
\author{A.~Bhattacharjee}
\affiliation{Max Planck/Princeton Center for Plasma Physics, Department of Astrophysical Sciences and Princeton Plasma Physics Laboratory, Princeton University, Princeton, NJ 08543, USA}

\begin{abstract}
We propose a new mechanism for turbulent mean-field dynamo in which the magnetic fluctuations resulting 
from a small-scale dynamo drive the generation of large-scale magnetic fields. This is in
stark contrast to the common idea that small-scale magnetic fields should be harmful 
to large-scale dynamo action. These dynamos occur in the presence of large-scale velocity shear and do not require net
helicity, resulting from off-diagonal components of the turbulent
resistivity tensor as the magnetic analogue of the ``shear-current'' effect.
Given the inevitable existence of non-helical
small-scale magnetic fields in turbulent plasmas, as well as the generic
nature of velocity shear, the suggested mechanism may help  explain
generation of large-scale magnetic fields across a wide range of astrophysical
objects. 
\end{abstract}
\maketitle

% INCLUDE URPIN REFERENCES AND SEND TO HIM
% INCLUDE SCHEKOCHIHIN SUGGESTIONS
% INCLUDE TOBIAS REFERENCES

%Astrophysical magnetic fields are observed to be well-correlated over
%length and time scales far exceeding that of the underlying fluid
%motions. Beautiful in its regularity, the 11-year solar cycle is the
%most well-known example of this behavior \cite{Hathaway:2010hz}. Such
%large-scale structure is puzzling given that strong magnetic fields
%are expected to emerge through the stretching and twisting of field
%lines by smaller scale turbulence, a process known as turbulent dynamo.
%If the fluid turbulence breaks statistical symmetry in some way (e.g.,
%through helicity), basic mean-field dynamo theory tells us to expect
%large-scale fields to develop due to the so-called $\alpha$-\emph{effect} \cite{Moffatt:1978tc,Krause:1980vr}.
%Here, the turbulence reacts to a large-scale field in such a way as
%to re-enforce it, causing instability. However, magnetic fields develop
%much faster at small scales than at large scales \cite{Boldyrev:2005ix}
%and can ``catastrophically quench'' the large-scale dynamo before
%it reaches observed amplitudes \cite{Kulsrud:1992ej,Vainshtein:1992fm}.
%In this work, we show that in non-helical systems with large-scale velocity shear,
%the small-scale dynamo can have the opposite effect, \emph{enhancing}
%the growth of large-scale fields. 

Astrophysical magnetic fields are observed to be well-correlated over
length and time scales far exceeding that of the underlying fluid
motions. Beautiful in its regularity, the 22-year solar cycle is the
most well-known example of this behavior \cite{Hathaway:2010hz}. Such
large-scale structure is puzzling given that strong magnetic fields
are expected to emerge through the stretching and twisting of field
lines by smaller scale turbulence.
As the primary theoretical framework to study such behavior, mean-field 
dynamo theory examines how large-scale magnetic fields develop
due to these small-scale turbulent motions. This splitting between scales
is captured by the mean-field average; the average of a fluctuating
quantity vanishes by definition ($\overline{ \bm{b}} =0$),
while the average of a large-scale field is itself ($\overline{\bm{B}} =\bm{B}$).
An average of the induction equation, which governs evolution of the
magnetic field within magnetohydrodynamics (MHD), leads to \cite{Moffatt:1978tc}
\begin{equation}
\partial_{t}\bm{B}=\nabla\times\left(\bm{U}\times\bm{B}\right)+\nabla\times\bm{\mathcal{E}}+\frac{1}{\mathrm{Rm}}\nabla^{2}\bm{B},\label{eq:MF eqn}
\end{equation}
where Rm is the magnetic Reynolds number, a dimensionless measure
of the plasma resistivity, and $\bm{U}$ and $\bm{B}$ are the large-scale
velocity and magnetic field. The \emph{electromotive force}, $\bm{\mathcal{E}}=\overline{\bm{u}\times\bm{b}} $,
is an average of the fluctuating fields ($\bm{u}$ and $\bm{b}$)
and responsible for dynamo action. In the early phases of a dynamo,
the mean fields can be considered a small perturbation to the underlying
turbulence. Combined with an assumption of scale-separation between
small-scale and mean fields, this allows a Taylor expansion \cite{Brandenburg:2005kla,Yokoi:2013di}
of $\bm{\mathcal{E}}$ in terms of $\bm{B}$, 
\begin{equation}
\bm{\mathcal{E}}=\alpha\circ\bm{B}+\beta\circ\nabla\bm{B}+\cdots,\label{eq:E expansion}
\end{equation}
where $\alpha$, $\beta$ are the tensorial transport coefficients, calculated from the
small-scale fields \cite{Krause:1980vr}. Since these depend on the
large-scale fields, a solution to Eq.~\eqref{eq:MF eqn} requires
knowledge of how $\bm{\mathcal{E}}$ changes with $\bm{B}$ (and possibly
$\bm{U}$), essentially a statistical closure for inhomogenous MHD. 

Historically, much work has focused on kinematic dynamo theory, in
which $\bm{u}$ is uninfluenced by the magnetic field \cite{Moffatt:1978tc,Brandenburg:2005kla}.
Kinematic theory predicts large-scale dynamo instability when the
fluid motions possess helicity, $\int \bm{u}\cdot \nabla \times \bm{u}\,d\bm{x} \neq 0$. However, the
applicability of such predictions has been called into question by
a number of authors \cite{Kulsrud:1992ej,Cattaneo:2009cx}. In particular,
above modest Reynolds numbers in both helical and non-helical flows,
the small-scale dynamo \cite{Schekochihin:2007fy} causes $\bm{b}$
to grow and saturate much more rapidly \cite{Boldyrev:2005ix} than
$\bm{B}$. This violates the kinematic assumption, both because $\bm{u}$
is altered before $\bm{B}$ grows significantly, and because a dynamically
important $\bm{b}$ exists independently of $\bm{B}$. The buildup
of small-scale fields is the origin of ``$\alpha$~quenching'', in which the mean field saturates well before reaching amplitudes
consistent with observation \cite{Gruzinov:1994ea,Bhattacharjee:1995ip,Blackman:2002fe,Hotta:2015bu}
due to the adverse influence of $\bm{b}$. 

In this letter we show that in turbulence with large-scale velocity shear, it is possible and realizable to have the small-scale dynamo \emph{enhance} the growth of the large-scale dynamo.
%Nonetheless, such quenching
%is somewhat specific to helical dynamos \cite{Park:2012eg}: is it feasible that the small-scale
%magnetic fluctuations could \emph{enhance} the growth of the large-scale
%dynamo in other situations? Here we show that this is possible and
%realizable in turbulence with large-scale velocity shear. 
We demonstrate
this both with statistical simulation \cite{Squire:2015fk}, in which
the effect is very clear but applies rigorously only at low Reynolds
numbers, and through calculation of transport coefficients from direct
numerical simulations (DNS). In addition, the existence of the effect has been confirmed analytically
using the second-order correlation approximation \cite{Analytic}, which
agrees with previous spectral $\tau$ approximation calculations \cite{Rogachevskii:2004cx}.

All computations presented here use the incompressible MHD model in the shearing box,
employing homogenous Cartesian geometry and periodic boundary conditions
in the shearing frame. With a mean flow $\bm{U}_{0}=-Sx\hat{\bm{y}}$ 
 imposed across the domain, this setup is designed
to represent a small ``patch'' of turbulent fluid in large-scale
velocity shear. 
We force with non-helical white-in-time noise at small scales and study the
generation of larger-scale magnetic fields, in a way similar to previous
authors \cite{Yousef:2008ix,Brandenburg:2008bc}. The mean-field average
is defined as an average over the horizontal ($x$ and $y$)
directions, such that the mean magnetic fields $\bm{B}$ depend only
on $z$. We also allow for system rotation through
a mean Coriolis force, since shear typically arises due
to differential rotation in astrophysical objects. The rotation $\bm{\Omega}$
is aligned with $\hat{\bm{z}}$ (antiparallel to $\nabla\times\bm U_0$), perpendicular to the flow $\bm{U}_{0}$.

For the chosen horizontal average, inserting Eq.~\eqref{eq:E expansion}
into \eqref{eq:MF eqn} gives 
\begin{gather}
\partial_{t}B_{x}=-\alpha_{yx}\partial_{z}B_{x}-\alpha_{yy}\partial_{z}B_{y}-\eta_{yx}\partial_{z}^{2}B_{y}+\eta_{ty}\partial_{z}^{2}B_{x}\nonumber \\
\partial_{t}B_{y}=-SB_{x}+\alpha_{xx}\partial_{z}B_{x}+\alpha_{xy}\partial_{z}B_{y}-\eta_{xy}\partial_{z}^{2}B_{x}+\eta_{tx}\partial_{z}^{2}B_{y},\label{eq:SC sa eqs}
\end{gather}
using velocity shear $\bm{U}_{0}$ but neglecting other mean velocities, and defining $\eta_{ti} \equiv \eta_{ii}+\mathrm{Rm}^{-1}$.
Here $\alpha_{ij}$ and $\eta_{ij}$ are the $\alpha$ effect and
turbulent resistivity tensors respectively, with the 4 components of 
$\eta_{ij}$ relatable to the $\beta_{ij3}$ elements of the full tensor [Eq.~\eqref{eq:E expansion}]. Due to homogeneity and
reflectional symmetry (vanishing net helicity), $\alpha_{ij}$ must vanish when averaged over
a suitably large time or number of realizations \cite{Brandenburg:2005kla}, and indeed our 
measurements confirm this.
There is no such constraint on $\eta_{ij}$, and $\eta_{yx}$ is very
important throughout this work due to its coupling with the shear.
In particular, neglecting fluctuations in $\alpha$ and assuming diagonal
resistivities are equal ($\eta_{tx}=\eta_{ty}=\eta_{t}$), the least
stable eigenmode of Eq.~\eqref{eq:SC sa eqs} for a mode of
vertical wavenumber $k$ grows at 
\begin{equation}
\gamma=k\left[\eta_{yx}(-S+k^{2}\eta_{xy})\right]^{1/2}-k^{2}\eta_{t}.\label{eq:gamma}
\end{equation}
Since $S\gg\eta_{ij}$, dynamo action is possible without an $\alpha$
effect if $\eta_{yx}<0$. 

Subsequent to early analytic work \cite{Urpin:1999wl,Urpin:2002ct,Rogachevskii:2003gg},
it was found kinematically that  $\eta_{yx}>0$ (at least at low Rm),
and several authors have thus concluded that a coherent shear dynamo
cannot explain observed field generation \cite{Radler:2006hy,Brandenburg:2008bc,SINGH:2011ho}.
Instead, a popular theory is that temporal fluctuations in $\alpha_{ij}$
cause an \emph{incoherent} mean-field dynamo. Importantly, in such
a dynamo, $\bm{B}\left(z,t\right)$ cannot have a constant phase in
time as it grows, since the average of $\bm{B}$ over an ensemble of realizations vanishes,
implying $\bm{B}$ must be uncorrelated with itself after $t\gtrsim(k^{2}\eta_{t})^{-1}$
\footnote{This condition may be altered if one considers the effects of magnetic helicity conservation, which may cause a local magnetic $\alpha$ effect as the large-scale field grows \cite{Brandenburg:2008bc}. However, this also causes coupling between different mean-field modes, and the detailed consequences of such an effect remain unclear.}.
While incoherent dynamos are possible in a variety of situations, here we argue for a different
situation -- magnetic fluctuations act to substantially decrease and potentially reverse the sign of $\eta_{yx}$,
causing the onset of a coherent large-scale dynamo that can overwhelm
the incoherent dynamo.

%%%%%%%%%%%%%%%%%%%%%%%%%%%%%%%%%
\begin{figure}
\begin{centering}
\includegraphics[width=1\linewidth]{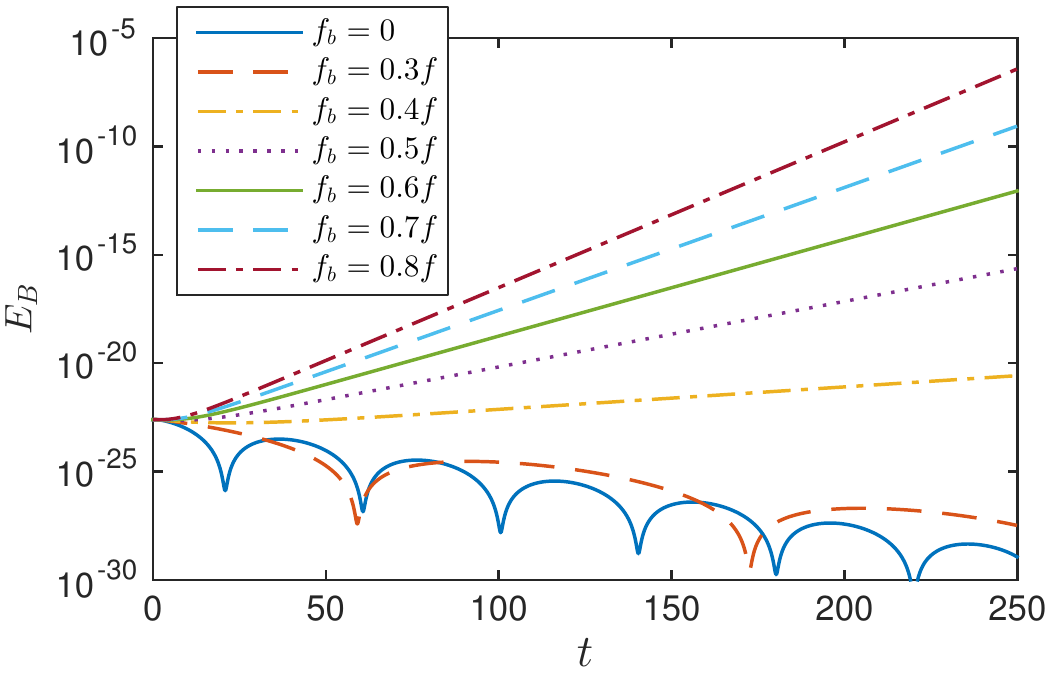}\caption{
Time development of the mean-field energy, $E_{B}=\int dz\,B^{2}/2$, in quasi-linear statistical simulation.
Small-scale fields are forced at $k_{f}=6\pi$, $\mathrm{Rm}=u_{rms}/\eta k_{f}\approx5$
(here $\eta$ is the resistivity, $\mathrm{Pm}=\mathrm{Rm}/\mathrm{Re}=1$),
$S=2$ and the box has dimensions $\left(L_{x},L_{y},L_{z}\right)=\left(1,1,4\right)$
with resolution $\left(28,28,128\right)$. As well as forcing
the momentum equation, the induction equation is forced to excite
homogenous magnetic fluctuations, emulating a small-scale
dynamo. Total forcing is kept constant in each simulation 
but the proportion of magnetic forcing ($f_{b}/f$) is increased from $0$ to $0.8$ \cite{Note2}.
As $f_{b}/f$ increases, the growth rate of the mean field increases
due to the change in sign and subsequent decrease of $\eta_{yx}$. \label{fig:CE2 sims}}

\par\end{centering}

\end{figure}
%%%%%%%%%%%%%%%%%%%%%%%%%%%%%%%%%%

Our first method illustrating this effect is quasi-linear statistical
simulation; see Ref.~\cite{LowRm} for further details. The method starts by forming equations for the fluctuating 
fields $\bm{u}$ and $\bm{b}$, and
linearizing these; i.e., neglecting fluctuation-fluctuation nonlinearities such as $\bm{u}\cdot \nabla \bm{u}$ and 
 $\bm{u}\cdot \nabla \bm{b}$.  One can then derive an equation
 for the fluctuation \emph{statistics}, $\mathcal{C} = \langle \chi_{i} \chi_{j} \rangle$ (where $\bm{\chi}=(\bm{u},\bm{b})$ and $\langle \cdot \rangle$ denotes the average over an
 ensemble of realizations), as a 
 function of $\bm{U}$ and $\bm{B}$
\cite{Farrell:2012jm,Tobias:2011cn,Squire:2015fk}.  
Finally, using $\mathcal{E}_{x}=\overline{\mathcal{C}_{26}-\mathcal{C}_{53}}$ and $\mathcal{E}_{y}=\overline{\mathcal{C}_{34}-\mathcal{C}_{61}}$, $\bm{\mathcal{E}}$ 
can be fed directly into Eq.~\eqref{eq:MF eqn},
resulting in a closed system of equations. Note that the method does not assume a scale-separation between 
 mean and fluctuating fields. Importantly, since the
statistics are calculated directly, an incoherent dynamo is not possible, and statistical simulation offers a direct
probe of the coherent effect.  The linearity of the fluctuation equations eliminates the
 small-scale dynamo; accordingly, to excite homogenous kinetic and magnetic fluctuations,
 both $\bm{u}$ and $\bm{b}$
are forced at small scales ($k_{f}=6\pi$) with 
the statistics of white-in-time noise.
The resulting MHD turbulent bath could
be considered as some approximation to kinetically forced turbulence 
after saturation of the small-scale dynamo.

To study the magnetically driven dynamo, we keep the total forcing
level constant, successively increasing the proportion of magnetic
forcing from purely kinetic \footnote{Specifically, for the 
forcing $\bm{f}_{i}$ (acting through
$\partial_{t}\bm{u}=\cdots + \bm{f}_{u}$ and $\partial_{t}\bm{b}=\cdots + \bm{f}_{b}$), 
we keep $\bm{f}=\bm{f}_{u}+\bm{f}_{b}$ constant (in the statistical sense) across all simulations in Fig.~\ref{fig:CE2 sims}}. Results with shear but no rotation are illustrated in Fig.~\ref{fig:CE2 sims}.
The presence of the magnetically driven dynamo is evident,
becoming slightly unstable when magnetic forcing accounts for $0.4$
of the total and increasing the growth rate thereafter. This sustained
period of exponential growth due to magnetic fluctuations is not possible
to see in DNS, since the mean field will immediately come into approximate
equipartition with the small-scale field due to the finite size of
the system.

%%%%%%%%%%%%%%%%%%%%%%%%%%%%%%%%%%%
\begin{figure}
\centering{}\includegraphics[width=1\linewidth]{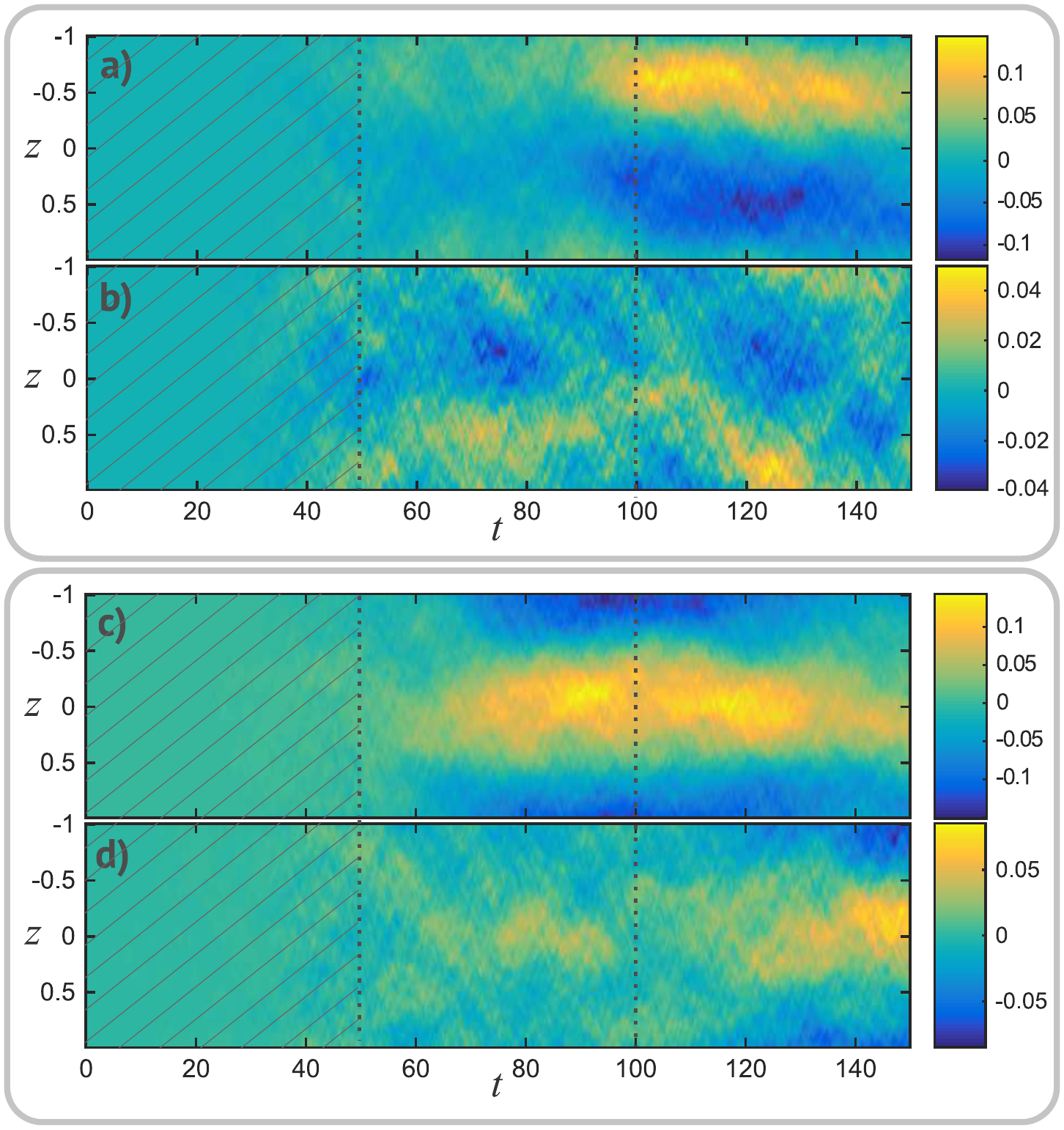}\caption{
Example spatiotemporal $B_{y}$ evolutions for non-rotating (a-b)
and Keplerian rotating (c-d) turbulence at $\mathrm{Rm}=u_{rms}/\eta k_{f}\approx15$
($k_{f}=6\pi,\,\eta=1/2000$, $\mathrm{Pm}=8$), $S=1$, in a box
of dimension $\left(1,4,2\right)$ with resolution $\left(64,128,128\right)$.
The first examples in each case {[}(a) and (c){]} show $B_{y}$ when
a coherent dynamo develops, while the second examples {[}(b) and (d){]}
illustrate the case when it is more incoherent. The main factors in
distinguishing these are the coherency in phase of $B_{y}$ over some
time period and the amplitude at saturation,
which is larger in the coherent cases. In general, the rotating simulations
are substantially more coherent. The hatched area illustrates the
region of small-scale dynamo growth. The fitting method used to compute
transport coefficients (see Fig.~\ref{fig:Transport coeffs}) is
applied between the dashed lines ($t=50\rightarrow100$).\label{fig:By examples} }
\end{figure}
%%%%%%%%%%%%%%%%%%%%%%%%%%%%%%%%%%%

The formal applicability of statistical simulation is limited to low
Rm due to the quasi-linear approximation. Our second method utilizes 
DNS, forced kinetically at small scales,  to show that magnetic fields arising consistently \emph{through
the small-scale dynamo} can drive a coherent large-scale dynamo.
To this end, we directly calculate transport coefficients from nonlinear
simulation before and after the saturation of the small-scale dynamo.
We use the incompressible MHD code SNOOPY \cite{Lesur:2007bh}, which uses 
the pseudo-spectral method and shearing periodic boundary conditions.
The chosen Reynolds numbers \cite{Yousef:2008ix} are moderate -- small enough
such that there is no self-sustaining turbulence in the absence of small-scale forcing (although
effects may be similar even when this occurs \cite{Lesur:2008cv}) --
and  ensembles of 100 simulations are run with shear $\bm{U}_{0}=-S x \hat{\bm{y}}$, 
both with and without Keplerian
rotation. At these parameters, the prevalence of the coherent large-scale
dynamo depends on the realization (see Fig.~\ref{fig:By examples}),
and it appears that the coherent effect cannot always overcome fluctuations
in $\bm{\mathcal{E}}$ immediately after small-scale saturation, although
the dynamo develops after a sufficiently long time {[}e.g.,
Fig.~\ref{fig:By examples}(d) near $t=150${]}. This behavior
seems generic when the coherent dynamo is close to its threshold for
excitation and we have observed similar structures when the induction
equation is driven directly at lower Rm \cite{LowRm}. Notwithstanding this variability
in the dynamo's qualitative behavior, measurement of the transport
coefficients illustrates that the $\eta_{yx}$ coefficient decreases
after the magnetic fluctuations reach approximate equipartition with
velocity fluctuations at small scales. 

%%%%%%%%%%%%%%%%%%%%%%%%%%%%%%
\begin{figure}
\begin{centering}
\includegraphics[width=1\linewidth]{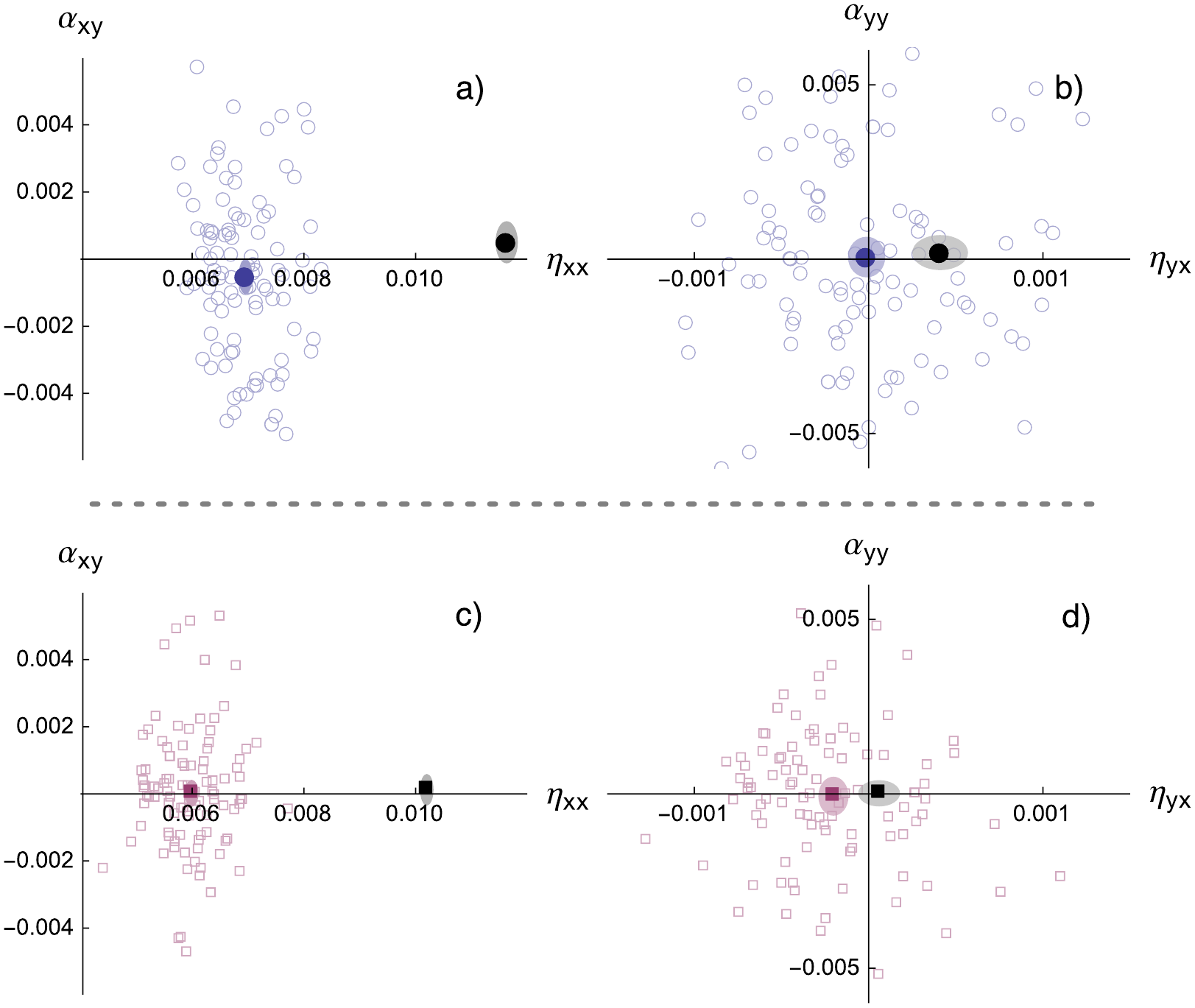}
\par\end{centering}
\centering{}\caption{
Measurements of the turbulent transport coefficients for 100 realizations
of the simulations at the same parameters as those in Fig.~\ref{fig:By examples};
(a) $\eta_{xx}$ and $\alpha_{xy}$ coefficients ($x$ and $y$ axes respectively), no rotation, (b) $\eta_{yx}$ and $\alpha_{yy}$ coefficients,
no rotation, (c) $\eta_{xx}$ and $\alpha_{xy}$ coefficients, rotating, (b) $\eta_{yx}$ and $\alpha_{yy}$
coefficients, rotating [see Eq.~\eqref{eq:SC sa eqs}]. Unfilled markers in each plot show coefficients
measured from each of the individual realizations, with mean values
displayed with solid markers and the shaded regions indicating error
in the mean (2 standard deviations). Black markers illustrate
the kinematic transport coefficients, with grey shaded regions indicating
the error. After saturation of the small-scale dynamo, we calculate $\alpha_{ij}(t)$ and 
$\eta_{ij}(t)$ by solving Eq.~\eqref{eq:E expansion} approximately (see text), taking the mean from
$t=50$ to $t=100$. This limited time window is chosen to avoid capturing
the saturation phase of the large-scale dynamo, since $\eta_{ij}$
is presumably modified in this phase. In both methods of computing
transport coefficients,
$\alpha$ coefficients are zero to within error
as expected, and the scatter between simulations is of a similar magnitude
to that of $\eta_{ij}$ if one accounts for their different units
(it is necessary to divide $\alpha$ by a characteristic $k$ value).
\label{fig:Transport coeffs}}
\end{figure}
%%%%%%%%%%%%%%%%%%%%%%%%%%%%%%%

\begin{figure}
\begin{centering}
\includegraphics[width=1\linewidth]{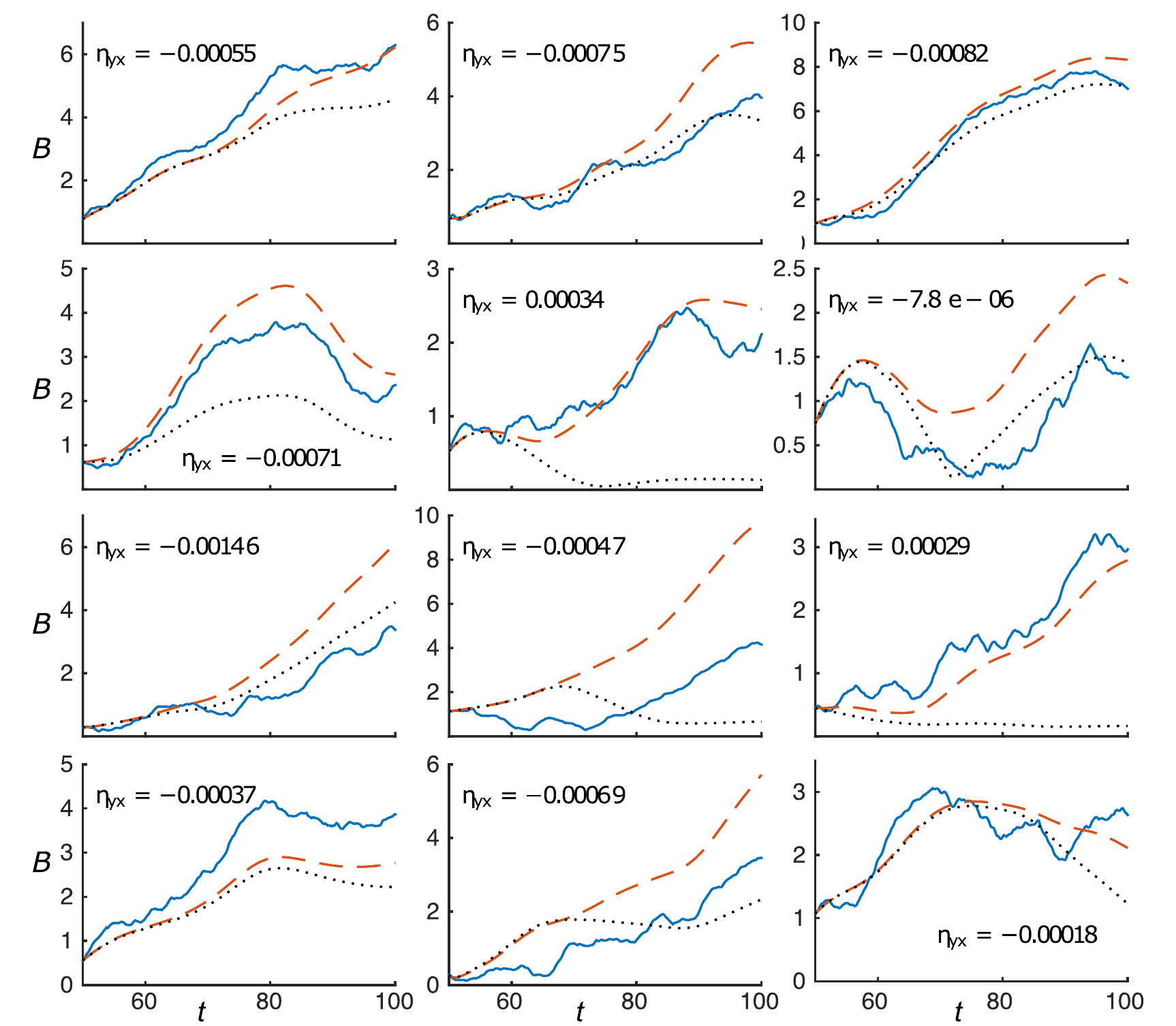}
\par\end{centering}
\centering{}\caption{
Evolution of the mean-field magnitude for a sample of the ensemble of rotating simulations 
(Figs.~\ref{fig:By examples}--\ref{fig:Transport coeffs}). Here $B$, the mean-field magnitude, is $(\vert\hat{B}_{x}^{1}\vert^{2}+\vert\hat{B}_{y}^{1}\vert^{2})^{1/2}$
where $\hat{B}_{i}^{1}$ is the largest scale Fourier mode of $B_{i}$. In each plot the solid blue curve 
shows data taken from the simulation. The dashed red curve shows the corresponding expected evolution, 
using the time-dependent calculated values of the transport coefficients, smoothed in time using a Gaussian filter of width $5$. 
Finally, the dotted black curve illustrates the expected evolution with all $\alpha$
coefficients artificially set to zero. The similarity between this evolution 
and that including $\alpha$ (dashed curve) illustrates that in many (but not all) cases
the dynamo is primary driven by $\eta_{yx}$. We list the measured mean of $\eta_{yx}$
in each plot to show that lower values  generally lead to stronger mean-field growth, as expected for a coherent dynamo. For reference, 
at the measured $\eta_{xx}\approx0.006$, the coherent dynamo is unstable below $\eta_{yx}=-0.00036$.
\label{fig:Realizations}}
\end{figure}

At low times, we use the test-field method to measure the kinematic
$\alpha$ and $\eta$, fixing the mean field and calculating $\bm{\mathcal{E}}$,
with no Lorentz force \cite{Brandenburg:2005kla,Brandenburg:2008bc}.
Since the small-scale dynamo grows quickly, test-fields are reset
every $t=5$. After small-scale saturation, standard test-field methods
are inapplicable \cite{Rheinhardt:2010do}. Instead, we extract $\bm{B}$
and $\bm{\mathcal{E}}$ simulation data and calculate $(\alpha_{ij}(t),\eta_{ij}(t))$
directly from Eq.~\eqref{eq:E expansion} by computing $\int dz\,\mathcal{E}_{i} \Theta$ for each of
$\Theta=(B_{x},B_{y},\partial_{z}B_{x},\partial_{z}B_{y})$ and solving the 
resulting matrix equations (in the least-squares sense) at each time point. 
This method is similar to that presented in Ref.~\cite{Brandenburg:2002cia}; however,
we additionally impose the constraints $\eta_{yy}(t)=\eta_{xx}(t)$, $\alpha_{xx}(t)=\alpha_{yy}(t)$ and $\alpha_{yx}(t)=\eta_{xy}(t)=0$. While these changes 
may appear to make the method less accurate, they in fact achieve the opposite by reducing 
the influence of $B_{x}$. This is necessary 
because $B_{x}$ has a high level of noise in comparison to its magnitude, and  because this noise is correlated with the noise in $B_{y}$ (due to $B_{x}$ directly driving $B_{y}$)
and $\mathcal{E}_{y}$ (through $\partial_{t}B_{x} = -\partial_{z}\mathcal{E}_{y}+\cdots$). These correlations are very 
harmful to the quality of the fit,  causing unphysical negative values for $\eta_{yy}$ \cite{Brandenburg:2002cia}, which also pollute measurement of other coefficients. 
It is straightforward to show that the systematic errors caused by our constraints on the transport 
coefficients are less than $1\%$ for the dynamos in Fig.~\ref{fig:By examples}, 
so long as $\eta_{xx}\approx\eta_{yy}$ when time averaged.
We have verified the method is accurate by comparison with the kinematic test-field method 
at low Rm, where there is no small-scale dynamo \cite{Yousef:2008ie,LowRm} and the rotation can be used to explore a range of $\eta_{yx}$. In addition, the measurements are independently 
verified in Fig.~\ref{fig:Realizations} below \footnote{Ideally, one would 
utilize the same method to measure transport coefficients before and after small-scale
saturation. However, the very fast growth of the small-scale 
field at early times renders the passive transport coefficient measurement
unfeasible, since the mean field is dominated by small-scale fields and
Eq.~\eqref{eq:E expansion} does not apply.}.
Due to the short time-window, measurements of the transport coefficients after small-scale saturation vary significantly between
realization, as should be expected from Fig.~\ref{fig:By examples}.
Nonetheless, an average over the ensemble of 100 simulations illustrates a statistically
significant change in $\eta_{yx}$ that is consistent with observed
behavior. 

%%%%%%%%%%%%%%%%%%%%%%%%%%%%%%
%%%%%%%%%%%%%%%%%%%%%%%%%%%%%%%

Results are illustrated in Fig.~\ref{fig:Transport coeffs}. In
the kinematic phase without rotation, we see $\eta_{yx}=\left(4.1\pm1.6\right)\times10^{-4}$
in qualitative agreement with previous studies \cite{Brandenburg:2008bc}.
With rotation $\eta_{yx}=\left(0.6\pm1.2\right)\times10^{-4}$,
consistent with a reduction in $\eta_{yx}$ due to the $\bm{\Omega}\times\bm{J}$
effect \cite{Krause:1980vr}. After saturation of the small-scale dynamo,
$\eta_{yx}=(-0.1\pm1.0)\times10^{-4}$ for the non-rotating case,
while $\eta_{yx}\approx-\left(2.0\pm0.8\right)\times10^{-4}$ in the
rotating case -- the same reduction in each to within error. Values
for the diagonal resistivity are smaller after saturation, as expected
since the velocity fluctuation energy decreases (by a factor $\sim1.4$).
The values of $(\eta_{xx},\eta_{yx})$ show that the dynamo
is slightly stable on average in the non-rotating case and marginal
in the rotating case. However, the coefficients vary significantly
between realizations, sometimes yielding larger growth rates, and
measurements match observed growth of the mean field for individual
realizations. We illustrate this in Fig.~\ref{fig:Realizations}, which demonstrates 
consistency between the measured transport coefficients and observed mean-field evolution
by solving Eq.~\eqref{eq:SC sa eqs} directly [using the time-dependent $\eta_{ij}$(t) and $\alpha_{ij}(t)$],
for a sample of the rotating simulations. In addition, by artificially removing $\alpha_{ij}(t)$, 
we illustrate that cases with more negative $\eta_{yx}$ are driven 
primarily by this, rather than a stochastic-$\alpha$ effect.
We thus conclude that small-scale
magnetic fluctuations act to \emph{decrease} $\eta_{yx}$, and that
in some realizations (or after a sufficiently long time period) a
coherent large-scale dynamo develops as a result.

To summarize, in this letter we have demonstrated that small-scale magnetic fluctuations, excited
by small-scale dynamo action, can drive large-scale magnetic field
generation. The mechanism is a magnetic analogue of the 
``shear-current'' effect \cite{Rogachevskii:2004cx,Rogachevskii:2003gg}, arising through the off-diagonal turbulent resistivity
in the presence of large-scale shear flow.
We have demonstrated its existence numerically using both DNS, 
with measurements of mean-field transport coefficients before and after small-scale dynamo 
saturation, and through quasi-linear
statistical simulation.

More work is needed to precisely assess regimes in which
the magnetically driven dynamo might dominate, as well as its behavior
at higher Reynolds numbers where self-sustained turbulence is possible \cite{Lesur:2008cv}.
Another interesting question regards whether a magnetic dynamo can
remain influential in the presence of net helicity and an $\alpha$ effect,
particularly as small-scale dynamo may be suppressed by shear \cite{Tobias:2014ek}.
While such questions may be difficult to answer definitively, the
generic presence of magnetic fluctuations in plasma turbulence gives
us some confidence that the proposed mechanism could cause large-scale
dynamo growth in the wide variety of astrophysical systems with velocity
shear.

\begin{acknowledgments}

This work was supported by a Procter Fellowship at Princeton University,
and the US Department of Energy Grant DE-AC02-09-CH11466. The authors
would like to thank A.~Schekochihin, J.~Krommes, and I. Rogachevskii
for enlightening discussion and useful suggestions. 
\end{acknowledgments}

%\bibliographystyle{apsrev4-1}
%\bibliography{fullbib}


\end{document}